\documentclass[conference]{IEEEtran}
\usepackage{cite}
\usepackage{amsmath,amssymb,amsfonts}
\usepackage{algorithmic}
\usepackage{graphicx}
\usepackage{textcomp}
\usepackage{xcolor}
\usepackage[hyphens]{url}
\usepackage{fancyhdr}
\usepackage[bookmarks=true,breaklinks=true,letterpaper=true,colorlinks,citecolor=blue,linkcolor=blue,urlcolor=blue]{hyperref}
\usepackage{adjustbox}
\usepackage{algorithm}
\usepackage{algorithmic}
\usepackage{amsmath}
\usepackage[flushleft]{threeparttable}
\usepackage{color}
\usepackage[normalem]{ulem}
\usepackage{adjustbox}
\usepackage{microtype}
\usepackage{xspace}
\usepackage{graphicx}
\usepackage{subcaption}
\usepackage{booktabs} 
\usepackage{amsmath}
\usepackage{bm}
\usepackage{multirow}
\usepackage{comment}

\usepackage{hyperref}
\usepackage{bm}
\RequirePackage[nameinlink]{cleveref} 
\crefname{chapter}{Chapter}{Chapters}
\crefname{section}{Section}{Sections}
\crefname{subsection}{Section}{Sections}
\crefname{equation}{Equation}{Equations}
\crefname{definition}{Definition}{Definitions}
\crefname{assumption}{Assumption}{Assumptions}
\crefname{theorem}{Theorem}{Theorems}
\crefname{figure}{Figure}{Figures}
\crefname{table}{Table}{Tables}
\crefname{algorithm}{Algorithm}{Algorithms}
\let\autoref\cref 
\newcommand{\Design}{\textbf{SAMT}\xspace}

\usepackage{pifont}

\title{Optimized Spatial Architecture Mapping Flow for Transformer Accelerators} 
\author{\IEEEauthorblockN{Haocheng Xu, Faraz Tahmasebi, Ye Qiao, Hongzheng Tian, Hyoukjun Kwon, Sitao Huang}
\IEEEauthorblockA{\textit{University of California, Irvine}\\
\{\textit{haochx5, tahmasef, yeq6, hyoukjun.kown, sitaoh}\}@uci.edu
}
}
\begin{document}
\maketitle
\pagestyle{plain}


\begin{abstract} 
Recent innovations in Transformer-based large language models have significantly advanced the field of general-purpose neural language understanding and generation. With billions of trainable parameters, deployment of these large models relies on high-performance hardware accelerators to efficiently deliver the required computation. Spatial architectures, such as TPUs, offer a promising solution to accelerating computation-intensive workloads. However, the design process for existing spatial architectures is predominantly manual, and it often involves time-consuming redesigns for new applications and new problem dimensions, which greatly limits the development of optimally designed accelerators for Transformer models. To address these challenges, we propose SAMT (Spatial Architecture Mapping for Transformers), a comprehensive framework designed to optimize the dataflow mapping of Transformer inference workloads onto spatial accelerators. We demonstrate the effectiveness of SAMT in improving the performance of spatial accelerators for Transformer models. We propose and leverage the dynamic operator fusion schemes for the Transformer models and co-search the optimal dataflow mapping strategies for spatial accelerators. SAMT significantly reduces inference latency by 12\% to 91\% and energy consumption by 3\% to 23\%  for evaluated Transformer models compared to traditional spatial accelerator designs among edge, mobile and cloud settings. 

\end{abstract}

\section{Introduction}
\label{sec:introduction}
\begin{figure}[t]
    \centering 
    \includegraphics[width=0.8\linewidth]{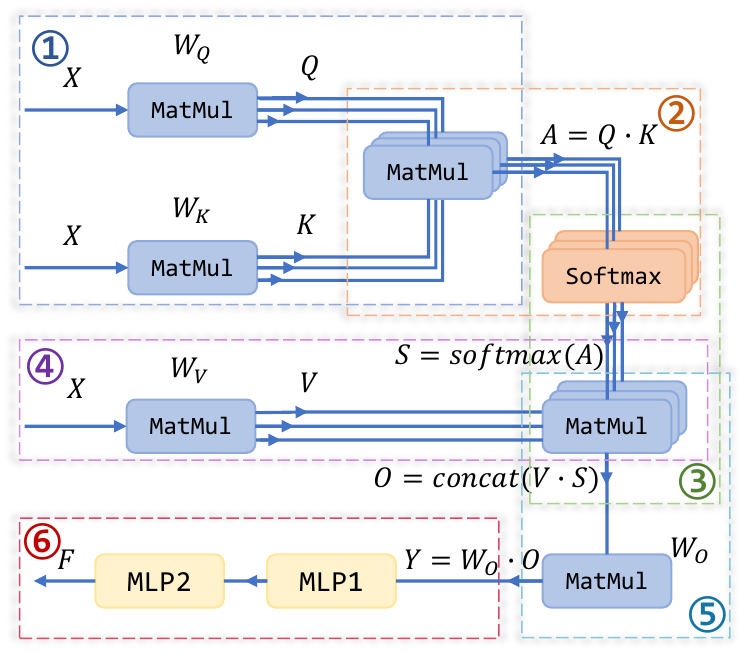}
    \caption{Major Computation Steps in a Transformer layer 
    }
    \label{fig:fusion_diagram}
\end{figure}

Machine learning (ML) workloads are gradually shifting from simple multilayer perceptrons (MLPs), feed-forward models, and convolutional neural networks (CNNs)  to large-scale Transformer-based models~\cite{vaswani2023attention}. Transformer-based models have delivered superior model performance (e.g., accuracy) compared to previous models such as CNNs in various application domains including computer vision~\cite{dosovitskiy2020image} and natural language processing (NLP) using large language models (LLMs) such as GPT-4~\cite{openai2023gpt4} and Llama2~\cite{touvron2023llama}. Their success was achieved with billions of trainable parameters with a large and complex computation graph in attention layers, which is the backbone of Transformer~\cite{vaswani2023attention}. Unlike CNNs which mainly consist of convolutions, self-attention blocks include multiple batched matrix multiplication (BMM) blocks and memory-bound operators (e.g., softmax)~\cite{kim2023full}, which leads to overall lower arithmetic intensity compared to CNNs. In addition, Transformer layers involve complex operator dependencies as shown in Fig. \ref{fig:fusion_diagram}. 
Such new characteristics of Transformer workload motivate specialized accelerator architectures for Transformers to exploit new optimization opportunities implied by the specific dependency patterns and low arithmetic intensity in Transformers. 

\begin{figure}
    \centering
    \includegraphics[width=\linewidth]{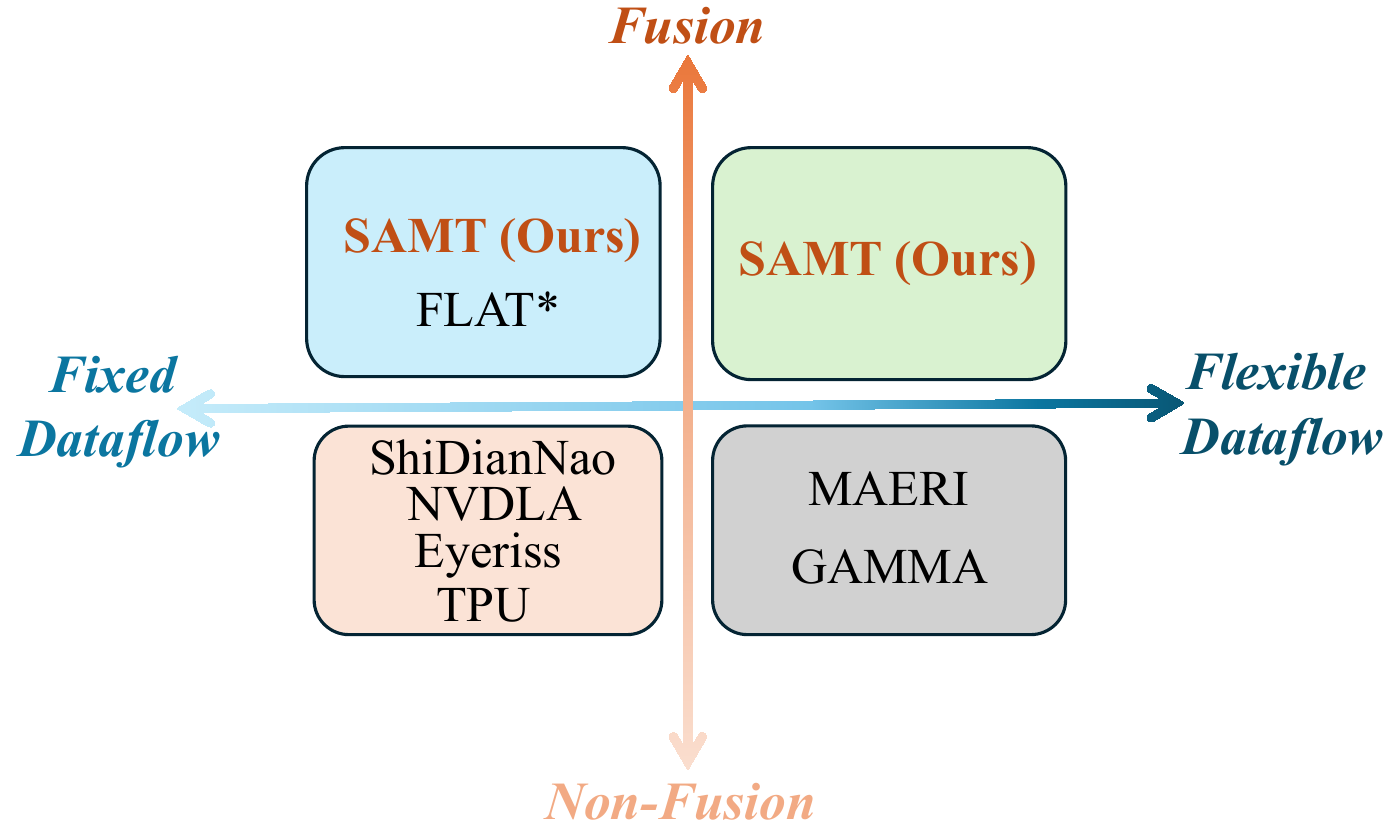}
    \caption{Lack of fusion exploration in spatial accelerators. Both traditional fixed dataflow accelerators (like ShiDianNao\cite{ShiDianNao}, NVDLA\cite{NVDLA}, Eyeriss\cite{eyeriss}, and TPU\cite{jouppi2023tpu}) and flexible dataflow accelerators (like MAERI\cite{MAERI} and GAMMA \cite{gamma}) did not consider operator fusion opportunities. *FLAT~\cite{kao2023flat} only considers one fusion while SAMT (ours) considers 64 fusion schemes. Details of accelerators without fusion can be found in Fig. \ref{fig:acc_type}. }
    \label{fig:position}
\end{figure}

\begin{figure*}[ht]
    \centering 
    \includegraphics[width=\linewidth]{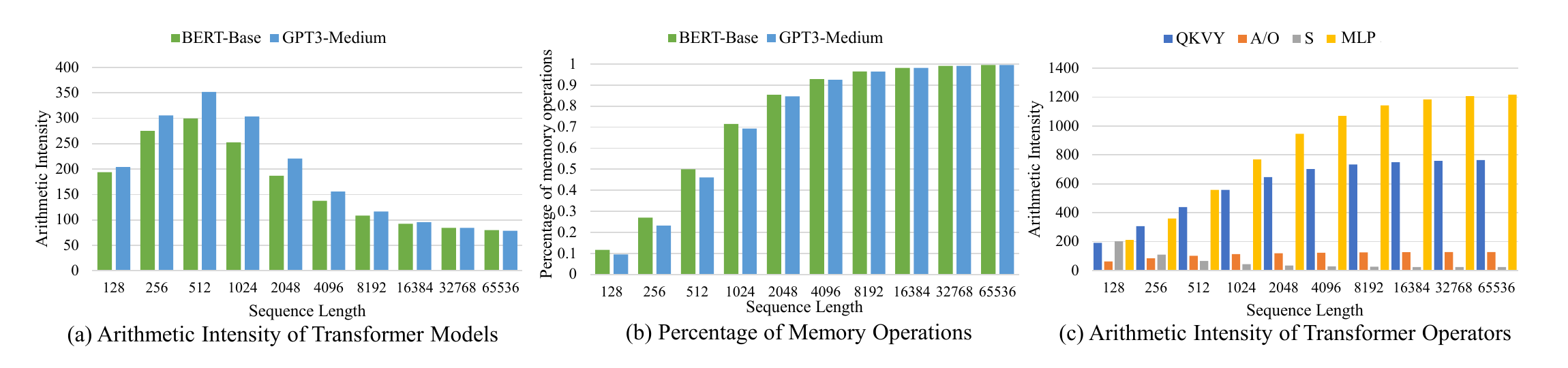}
    \caption{(a) Arithmetic intensity with different sequence length $l$ for BERT-Base model (embedding size $d$ = 768 and number of heads $n_h$ = 12) and GPT3-Medium model's prefilling stage ($d$ = 768 and number of heads $n_h$ = 12) (b) The percentage of the memory operations for the operations (for example, $A$ and $S$ in Fig. \ref{fig:fusion_diagram}) that scale with the sequence length. (c) Arithmetic intensity for different operators within the same transformer model. }
    \label{fig:Arithmetic Intensity}
\end{figure*}

In addition, recent trend in the usage of large language models such as ChatGPT~\cite{openai2023gpt4} indicate that the input sequence length is sharply increasing for complex tasks such as document summarization~\cite{Unlimiformer}, presentation generation~\cite{zhang2024genserp}, and video generation~\cite{sora}.
For example, the upper limit of the number of input tokens for GPT-35-Turbo is 4K (4096), GPT-4 is 8K and GPT-4-32K is 32K~\cite{Token}. Such a trend is imposing a major challenge to the hardware since longer sequence length results in lower arithmetic intensity, as we show in Fig. \ref{fig:Arithmetic Intensity}. Although the arithmetic intensity of the Transformer model increases until the sequence length of 512, it starts decreasing sharply beyond 512 tokens, which introduces a severe memory bottleneck problem for recent use cases with 4K or longer input sequences. 

As a solution for such a low arithmetic intensity challenge, operator fusion, which merges sequential operations into a single composite operation and reduces unnecessary global memory and off-chip memory access between sequential operators, has been explored~\cite{alwani2016fused, gao2019tangram, cai2023inter, symons2023stream}. However, most previous works either focus on GPUs~\cite{dao2022flashattention,dao2023flashattention2} or CNNs~\cite{alwani2016fused, gao2019tangram}. Some works covered Transformer targeting accelerators, but the operator fusion search space is limited to batch dimension~\cite{cai2023inter} or a subset of operators in Transformer blocks~\cite{kao2023flat}. Recent advances in machine learning algorithms enabled full integer arithmetic in Transformer blocks~\cite{xiao2024smoothquant,kim2021ibert,dettmers2022llmint8} and revealed more unexplored operator fusion opportunities in Transformer blocks, which motivates a comprehensive study like this work.

However, exploring the new operator fusion search space is not a trivial optimization problem. Not only does it add more dimensions of complexity to the dataflow design space, but it explicitly depends on the underlying hardware accelerator and resources. Spatial accelerators often consist of thousands of processing elements (\emph{PEs})~\cite{jouppi2023tpu} with kilobytes of local (\emph{S1}) and megabytes of shared (\emph{S2}) scratchpad memories for data reuse and a customized network-on-chip (\emph{NoC}) for interconnection. Due to the increasing scale of the hardware, the hardware design space is large, which is easily in a scale of $O(10^{12})$ for a fully flexible accelerator based on MAERI \cite{MAERI} architecture estimated for 1.72 mm\textsuperscript{2} of chip area using 15 nm technology. Moreover, the number of available dataflow mappings on the hardware outnumbers the hardware design space due to the various loop ordering, tiling, and temporal scheduling options, particularly with the large tensor dimensions in the Transformer-based models. For example, mapping the projection layer of Llama2-13B \cite{touvron2023llama} where there are 40 attention heads, each with a dimensionality of 128, embedding size of 5120, and the input sequence length of 1024, onto a 16$\times$16 PE array with 128 KB of on-chip scratchpad memory results in $O(10^{11})$ potential dataflow mappings. The interplay between hardware design and dataflow mapping is crucial for optimal accelerator performance, yet co-optimizing these aspects remains a significant challenge. The problem is even more challenging when considering the operator fusion with the complex dependency in Transformer models together. 

Therefore, to address these challenges,
we propose SAMT (Spatial Architecture Mapping for Transformers), as shown in Fig. \ref{fig:framework}, a comprehensive framework designed to optimize the dataflow mapping of Transformer inference workloads onto spatial accelerators, with operator fusion optimizations. Unlike previous works, SAMT considers the full spectrum of operator fusion possibilities, accommodating various hardware configurations, including the number of PEs, local (\emph{S1}) and shared (\emph{S2}) scratchpad memory sizes, and \emph{NoC} bandwidth. 
Our proposed SAMT flow models all 
the possible coarse-grained operator fusion opportunities when mapping Transformer workloads to spatial accelerators, and explore the combined design space of operator fusion and dataflow mapping. To the best of our knowledge, SAMT is the first work that optimizes the spatial architecture mapping considering all possible operator fusion opportunities in a Transformer block.

The major contributions of this work can be summarized as follows. 
\begin{itemize}
  {\item We develop a fused dataflow mapping optimization framework, \textit{\textbf{SAMT}}, which maps Transformer inference workloads to spatial accelerators with a fixed or flexible dataflow.}
  {\item We thoroughly explore all the possible operator fusion opportunities in Transformer blocks and show new optimization opportunities in the expanded fusion space, which on average lead to 21\% and 15\% lower latency and energy consumption, respectively, compared to state-of-the-art works with operator fusion.}
  {\item We develop an operator fusion explorer (\textit{\textbf{OFE}}), which automatically identifies the optimal fusion scheme for a given Transformer inference workload and a target spatial accelerator.}
  {\item We extend an analytical cost model, MAESTRO, to support operator fusion (\textit{\textbf{MAESTRO\_ FUSION}}) that can evaluate the performance of fused and non-fused operators running on spatial accelerators.}
  {\item We propose a mapping space explorer (\textit{\textbf{MSE}}), a genetic algorithm based mapper that identifies Pareto-optimal mapping strategies that support both fixed and flexible dataflow accelerators.}

\end{itemize}

\section{Background}
\label{sec:background}

\begin{figure}[ht]
    \centering 
    \includegraphics[width=0.9\columnwidth]{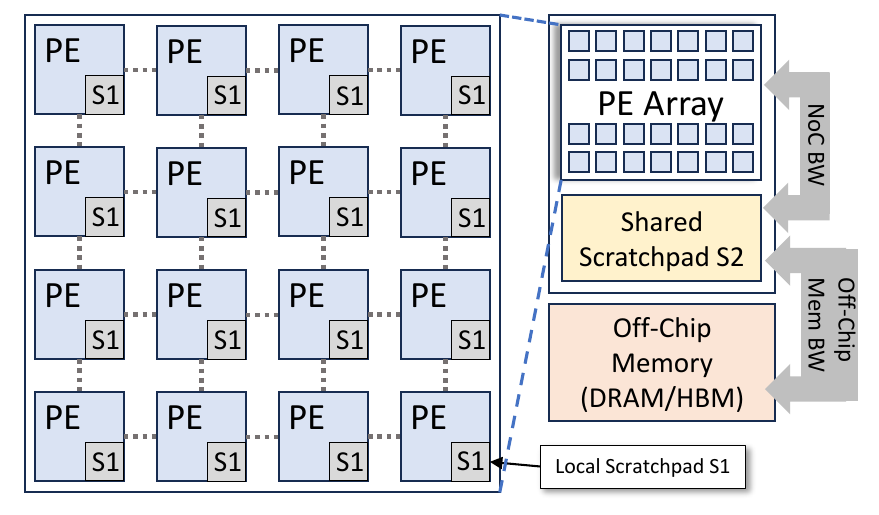}
    \caption{Abstraction of spatial accelerator architecture}
    \label{fig:spatial arch}
\end{figure}
\begin{figure}[ht]
    \centering 
    \includegraphics[width=1\columnwidth]{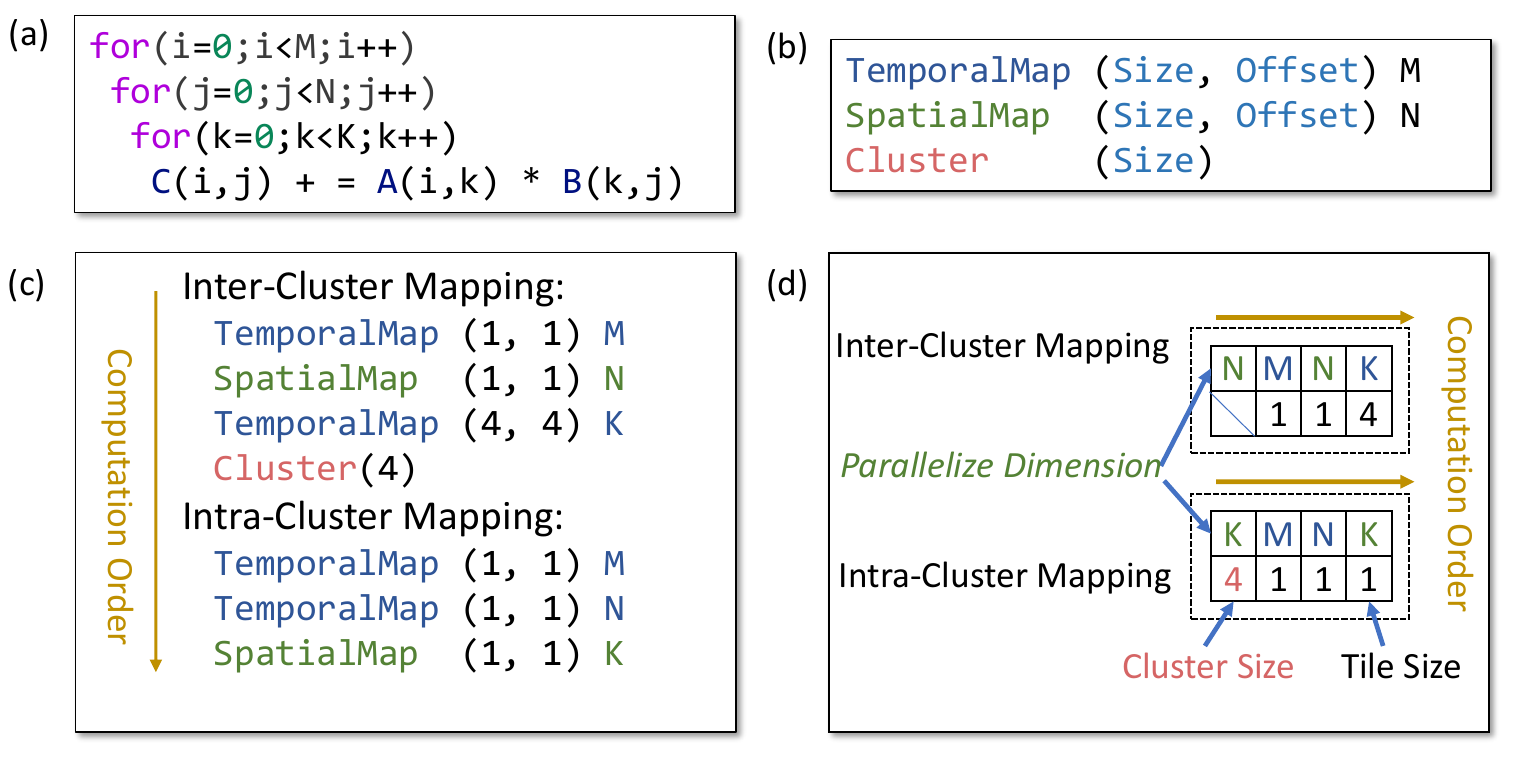}
    \caption{Example of mapping in MAESTRO\_FUSION}
    \label{fig:mapping_description}
\end{figure}
To better understand the need and challenges for accelerating Transformer models with spatial accelerators, we elaborate on the details from model, mapping, and hardware perspectives in this section.


\subsection{Spatial Accelerators and Dataflow Mapping }

\begin{figure*}[ht]
    \centering 
    \includegraphics[width=\linewidth]{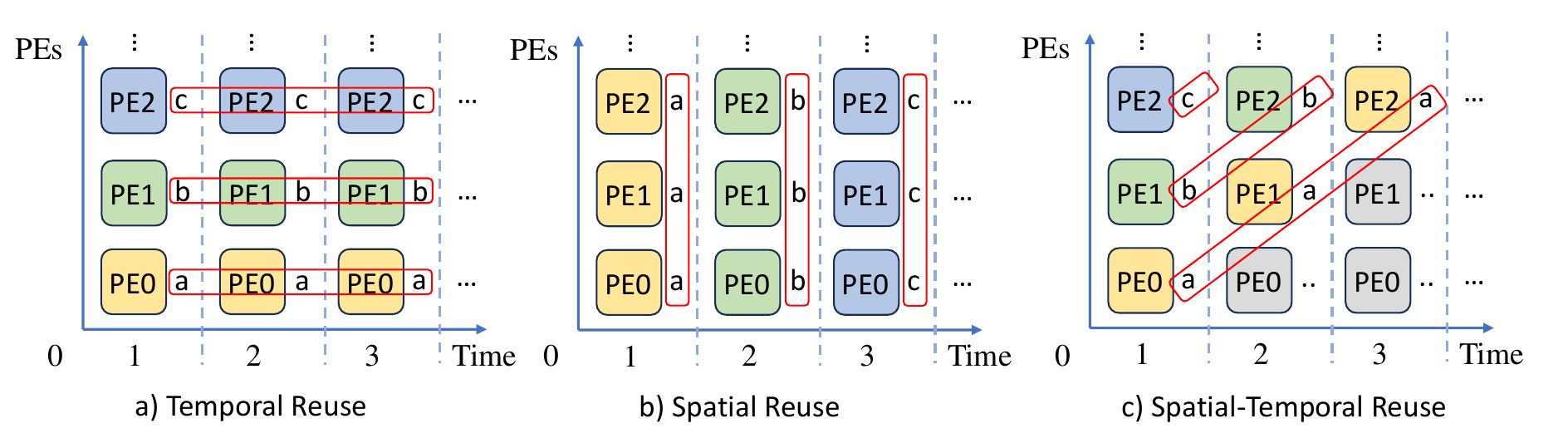}
    \caption{Understanding data reuse in spatial architecture: \textbf{Temporal} reuse happens when PEs access the same data across time via on-chip scratchpads or buffers. In Fig. \ref{fig:data_reuse}(a), PE0, PE1, and PE2 stored the same data $a$, $b$, and $c$ for the timestamps 1, 2, 3, and so on. \textbf{Spatial} reuse occurs when multiple PEs access the same data at the same time via multicasting. In Fig. \ref{fig:data_reuse}(b), all the PEs receive the same data $a$ at timestamp 1, data $b$ at timestamp 2, and data $c$ at timestamp 3. \textbf{Spatial-temporal} reuse forwards the data to the adjacent PEs in a skewed manner via neighbor-to-neighbor connections with forward buffers. In Fig. \ref{fig:data_reuse}(c), PE0 sends the data $a$ to adjacent PE1 from timestamp 1 to 2 and PE1 sends the data to PE2 from timestamp 2 to 3.}
     \vspace{-4mm}
    \label{fig:data_reuse}
\end{figure*}

As shown in Fig. \ref{fig:spatial arch}, a spatial accelerator architecture consists of a PE (processing element) array, local scratchpad memory $S1$ within each PE, a shared scratchpad memory $S2$, and NoCs (network-on-chips) that interconnect PE array and shared scratchpad memory $S2$. PEs serve as computation units within accelerators, containing ALUs (arithmetic logic units) or MAC (multiply-accumulate units) with a local scratchpad $S1$. Similar to cache memory, scratchpad memory is used to reduce the number of accesses for off-chip memory $S3$ and it provides the flexibility for the hardware designers to customize data layout and the process of reading/writing the data. The spatial accelerators have different inter-PE communication patterns as well. With the flexibility of spatial accelerators, we can further explore data reuse opportunities. Most layers in the transformer involve tensor-tensor MAC operations and provide a huge space for data reuse opportunities along different dimensions of the tensors~\cite{dataorchestration}. These data reuse opportunities could be \textbf{temporal}, \textbf{spatial}, or \textbf{temporal-spatial}. 
As we can tell from the Fig. \ref{fig:data_reuse}, PEs in spatial accelerators can communicate directly using special NoC and provide three types of data reuse, $temporal$ reuse, $spatial$ reuse, and $spatial-temporal$ reuse, via multicasting, broadcasting, forwarding, and reduction as part of the NoC. 


We define dataflow as the mechanism to leverage the maximum data reuse opportunity in domain-specific accelerators. Dataflow contains three typical mechanisms: \textbf{parallelization strategy}, \textbf{loop order}, and \textbf{tiling strategy}. 
The parallelization strategy specifies spatial and temporal mappings for each dimension. For example, setting a chunk size of one in a temporal dimension uniformly distributes data indices across all Processing Elements (PEs), facilitating spatial data reuse within the same timestep. The sequence of parallel and temporal loops in a loop nest directs data movement and impacts data mapping to PEs over time. Alterations in loop order can drastically change data's stationary behavior. Tiling subdivides mapping dimensions into smaller segments, increasing mapping flexibility and overcoming the limitations of traditional loop nests. Unlike standard loops, which complete all inner loops before progressing, tiling allows for dynamic loop execution. This can reduce peak memory demands per PE and enable strategic exploitation of partial temporal reuse opportunities.


In Fig. \ref{fig:mapping_description}, we provide an example of how to map a GEMM with fixed dataflow using dataflow directives in \textbf{MAESTR\_FUSION}.  Fig. \ref{fig:mapping_description} (a) describes a matrix multiplication between matrix $A$ ($M \times K$) and matrix $B$ ($K \times N$) and produce a matrix
$C =A\otimes B$ with dimension $M$ = 3, $N$ = 3, $K$ = 3. Fig. \ref{fig:mapping_description} (b) shows how we can use dataflow directives to describe the mapping of GEMM. \textbf{TemporalMap} indicates that data will change over time and remain static over PEs. \textbf{SpatialMap} implies data will change over space (across PEs). \textbf{Cluster} controls how many PEs will be grouped into a cluster. \textbf{Size} means mapping size, which tells us how much data is mapped onto each PE or how many PEs are mapped onto each cluster. \textbf{Offset} means Temporal Offset, which means the mapping update over time, or Spatial Offset, which means spatial mapping across the PEs with offset. Fig. \ref{fig:mapping_description} (c) shows the dataflow directives for this GEMM in (a). We used three directives to describe the GEMM: Inter-Cluster Mapping, Cluster, and Intra-Cluster Mapping.

\textbf{Cluster}: it will divide the entire PE array into different clusters, each cluster has an equal amount of PEs. For example, we can assume the current accelerator has 6 PEs in total, and Fig. \ref{fig: different mappings} (c) and (d) indicate that these 6 PEs are divided into 2 groups and each group has 3 PEs. This enables exploration in a 2D PE array by spatially mapping two dimensions of the GEMM. \textbf{Intra-Cluster Mapping}: it describes the details of mapping within the cluster. in Fig. \ref{fig: different mappings} (c), the $K$ dimension is spatially mapped (data are different across PEs) while the $M$ and $N$ dimensions are temporally mapped (data only change with time). With $Size = 1$ and $Offset = 1$, each PE will only receive one unique element from matrices $A$ and $B$ since $M$ and $N$ stay the same. Each cluster will compute one element of matrix $C$. \textbf{Inter-Cluster Mapping}: it describes a similar mapping process but across the clusters. In Fig. \ref{fig: different mappings} (c), the $N$ dimension is spatially mapped, and $M$ and $K$ dimensions are temporally mapped. This implies the elements from the matrix $B$ will be distributed across clusters while the elements from the matrix $A$ remain the same. The $Size$ and $Offset$ for K is 3 means each cluster receives 3 elements from each matrix. The computation order depends on the directives described in Fig. \ref{fig:mapping_description} (c) and (d). 


\subsection{Computation in Transformer and the low arithmetic intensity challenge}
\label{subsec:transformer computation}
Transformer models typically consist of two main blocks, which are usually referred to as the \emph{attention} block and the \emph{feed-forward} module, each of which is followed by a LayerNorm and a residual connection \cite{vaswani2023attention}.

As illustrated in Fig. \ref{fig:fusion_diagram}, attention blocks and feed-forward module first project the sequence input $X$ of dimension $d \times l$ ($d$ and $l$ represent the embedding size and sequence length, respectively) by different weight matrices $W_Q$, $W_K$, $W_V$ of dimension $d \times d$ to get the so-called \emph{query} $Q$, \emph{key} $K$, and \emph{value} $V$ matrices of dimension $d \times l$. The $Q$ matrix and transposed $K$ matrix will yield an activation matrix $A$ of size $l \times l$. Then attention score $S$ will be calculated based on the softmax value of the activation matrix $A$. To get the output of the attention block, it will multiply the value matrix $V$ with the attention score $S$, resulting in a matrix $O$ with a dimension of $d \times l$. Finally, to get the output $Y$ of the entire attention block, it projects the matrix $O$ by one more weight matrix $W_O$. After the attention block, the intermediate tensor will be passed into two consecutive linear projection layers $MLP$ with a given hidden dimension $d_{FFN}$ plus two GELU (Gaussian Error Linear Unit) non-linear activation layers in between. 



To understand the reason behind the limits of tokens in the recent transformer models as we discussed in \ref{sec:introduction}, we profiled the arithmetic intensity $I$, which is calculated by dividing FLOPs (the number of floating-point operations) by MOPs (the number of memory access counts) (assuming that each memory access will fetch one-byte data):

\begin{equation}
    I = \frac{FLOPs}{MOPs}
\end{equation}

As shown in Fig. \ref{fig:Arithmetic Intensity}(a), the arithmetic intensity first increases then decreases after reaching sequence length $l$ = 512 for both the BERT-Base and the GPT3-Medium models. The main reason for this drop in arithmetic intensity is that some operations' (like $A = Q \otimes K$ and $S = Softmax (A)$) memory access count will scale up quadratically with the sequence lengths $l$ while keeping the number of floating point operations at low level. In Fig. \ref{fig:Arithmetic Intensity}(b), we can see that those operations will quickly dominate in memory access count in the entire transformer model as sequence length $l$ increases. 

As shown in Fig. \ref{fig:Arithmetic Intensity}(c), we can find some operators have low arithmetic intensity. As sequence length increases, Operator $QKVY$ (projection) and $MLP$'s (feed-forward) arithmetic intensity increases accordingly while the other two type operators $A/O$ and $S$'s arithmetic intensity stay steady or even decreases. This further proves that those memory operations will become a bottleneck when sequence length increases in the transformer model.

\subsection{Lack of fusion exploration in flexible dataflow accelerator }

There are several hardware-aware compiler frameworks \cite{chen2018tvm,xla} for fusion optimization. However, they can not directly target flexible dataflow spatial accelerators. FLAT\cite{kao2023flat} tries to target fixed dataflow accelerators and perform inter-layer fusion; however, it is limited to activation-to-activation fusion and does not fully explore all the fusion schemes. There are two main reasons why fusion exploration has been limited on spatial accelerators: 1) Most spatial accelerators suffered from model performance due to integer arithmetic; 2) Non-linear operators in transformer models like softmax and layer norm do not naturally support tiling due to the nature that it requires global views of the data(for example, softmax needs the maximum of the row and summation of the row). That means if we fused the naive safety softmax with other adjacent operators, a very large local memory will be needed to store all the data for softmax calculation, which is not practical in the real world.

Luckily, a recent study \cite{xiao2024smoothquant} enables lossless 8-bit weight and activation quantization for LLMs which further expands the usage of spatial accelerators in transformer models. The difficulty of fusing non-linear operators has also been addressed by the recent GPU work FlashAttention \cite{dao2022flashattention,dao2023flashattention2}.

To the best of our knowledge, there is no prior work that removes all limitations on the fusion exploration of dataflow accelerators. Despite increasing the design search space, fusion exploration brings great optimization opportunities. The goal of this work is to introduce such a framework that finds the best fusion scheme with arbitrary dataflow to be exploited by a flexible hardware accelerator, where it introduces a novel approach to computing attention in Transformer models, particularly emphasizing the importance of operation fusion to handle the intensive memory requirements and computational challenges of one-head attention mechanisms.

\section{\Design Spatial Accelerator Mapping For Transformer}
\label{sec:methodology}

\begin{figure*}[ht]
    \centering 
    \includegraphics[width=\linewidth]{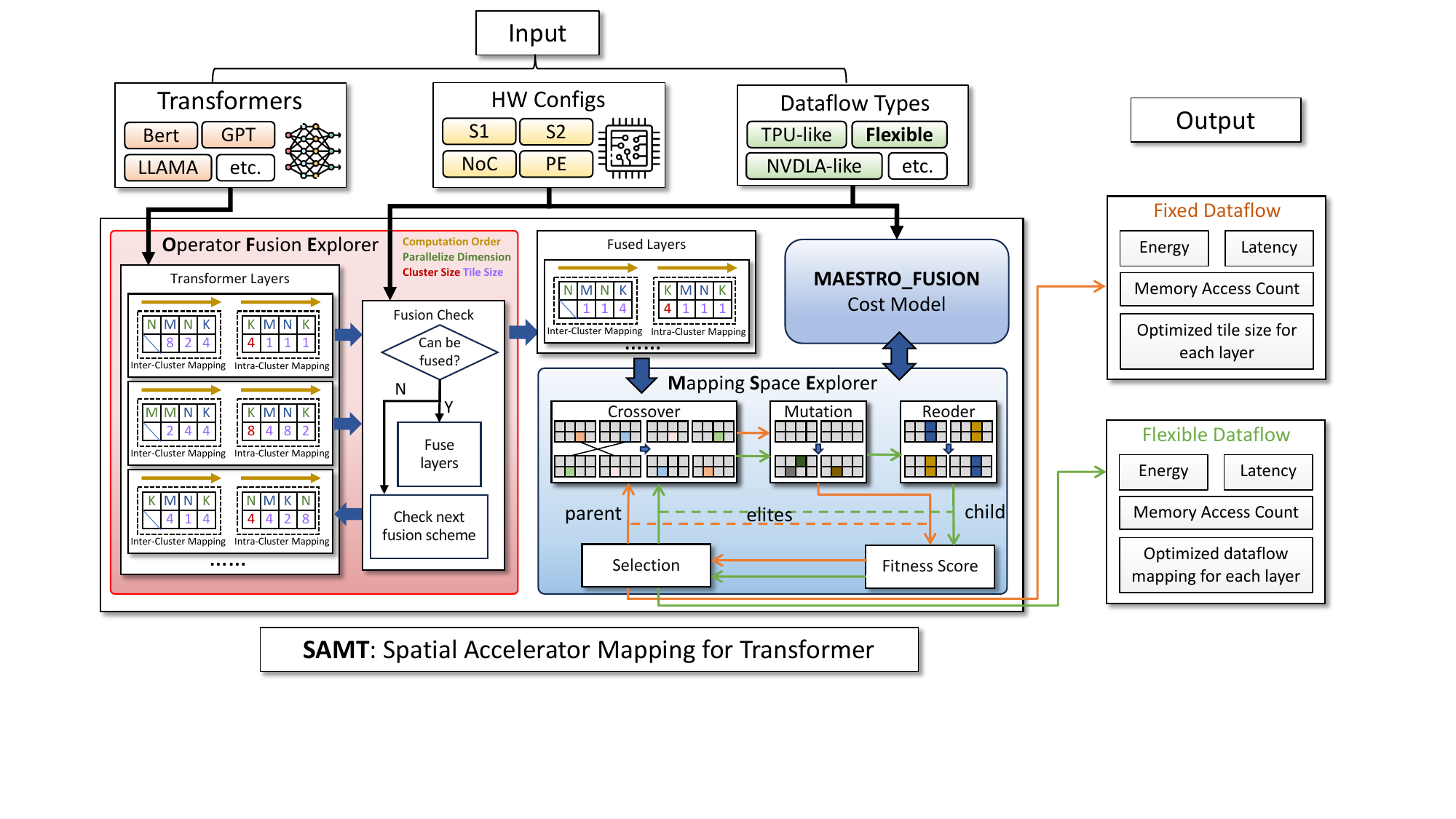}
    \caption{SAMT: Spatial Transformer Architecture Mapping Framework }
    \label{fig:framework}
\end{figure*}

Based on the motivation toward thorough exploration of fused mappings in Transformer layers enabled by recent integer arithmetic algorithms discussed in Section~\ref{sec:background}, we develop a fused mapping optimization framework, \Design, as illustrated in Fig. \ref{fig:framework}. \Design receives workload description, target accelerator hardware design parameters, and dataflow configuration (fixed or flexible) as inputs. Using those inputs, \Design reports optimized fused mapping and corresponding latency, energy, and memory access count. \Design consists of OFE, MSE, and MAESTRO\_ FUSION. We discuss each component in \Design in detail.


\subsection{OFE: Operator Fusion Explorer}
\label{subsec:ofe}
As we discussed in \ref{sec:background}, the main source of latency and energy benefits from operator fusion is based on the conversion of costly S2/DRAM accesses into efficient inter-PE communication. However, the amount of benefits are dependent on the model parameters (embedding size, sequence length, and hidden dimension), as listed in TABLE.\ref{tab:op_fuse}. Also, S2 memory size is another major consideration because it constraints available fusions (one or combination of what is listed in TABLE.\ref{tab:op_fuse}. Therefore, we develop \textbf{OFE}, Operator Fusion Explore, which systematically explores the complex design space of the fused dataflow and accelerator hardware parameters.

The design space for finding the optimal dataflow mapping given a fused or non-fused transformer model is large. As mentioned in \ref{sec:introduction}, mapping the projection layer of Llama2-13B \cite{touvron2023llama} can result in $O(10^{11})$ different dataflow mappings. Inspired by previous works' effectiveness of genetic algorithms for exploring such complex mapping space~\cite{gamma, kao2022magma, suda2016throughput, vasilache2018tensor}, we also adopt a genetic algorithm (Algorithm \ref{alg:SAMT_kernel_fusion}) with an extended search space with fusion.
For that, we develop new gene encoding for each operator into a 4 by 2 dimension of the genome as illustrated in Fig. \ref{fig:mapping_description}(d). Here we define genome as a design choice of the entire mapping space for dataflow. As shown in Fig.\ref{fig:mapping_description}(d), there are two levels of mapping corresponding to the inter-cluster and intra-cluster mapping. In the intra-cluster mapping, the first column describes the dimension ($K$) to parallelize and the following three columns describe the computation order and the tiling size for each TemporalMap or SpatialMap. In the inter-cluster mapping, the same terminology applies here. The first column in the inter-cluster mapping describes the dimension ($N$) to map spatially and the following three determine the order of computation and the corresponding tiling size. 

\begin{figure}[ht]
    \centering 
    \includegraphics[width=0.6\columnwidth]{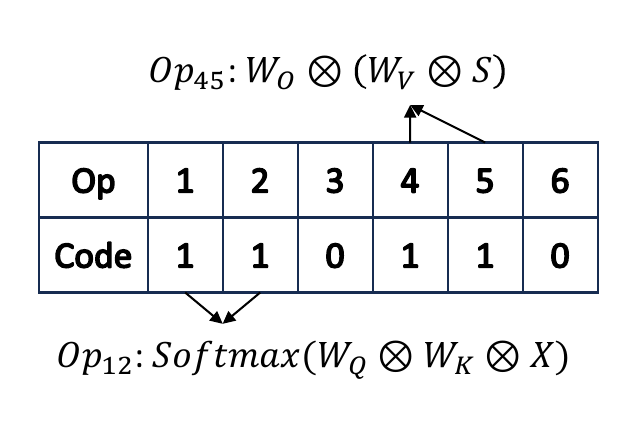}
    \caption{Example of new fused operators in \textbf{OFE}}
    \label{fig:new_op}
\end{figure}

We define the fused operator in TABLE \ref{tab:op_fuse} as our smallest primitives and encode the entire fusion scheme into a binary code. For example, as shown in Fig.\ref{fig:new_op}, fusion scheme $110110$ means we are using two newly fused operators, $Op_{12}$ and $Op_{45}$. $Op_{12}$ means fusing the original two fused operators $Op1$ and $Op2$. $Op_{12}$ needs to read matrix $W_Q$, $W_K$, $X$ from the off-chip memory $S3$ and will compute $S = Softmax(W_Q \otimes W_K \otimes X)$, following by writing $S$ back to the off-chip memory $S3$. Compared to the non-fused version, it will reduce the entire memory footprint by $5dl+2l^2$ with larger $S2$ requirements. Similarly, $Op_{45}$ will reduce the entire memory footprint by $3dl$. \textbf{OFE} will generate all the fusion schemes to be further evaluated. 

We can fuse some adjacent operators to reduce the intermediate tensor's memory footprint and increase the data reuse. The new fused operators are defined in TABLE \ref{tab:op_fuse} and can be visualized in Fig. \ref{fig:fusion_diagram}. As illustrated in Fig. \ref{fig:fusion_diagram}, the original computation to get the intermediate results $A$ is: 1) loading input $X$ twice and weight matrices $W_Q$ and $W_K$ with the input tensor memory footprint of $2dl + 2d^2$; 2) multiplying the $X$ with $W_Q$ and $W_K$ separately, and storing the intermediate results $Q$ and $K$ with the memory footprint of $2dl$; 3)loading the intermediate results $Q$ and $K$ with the same footprint; 4) multiplying $Q$ and $K$ to get the intermediate results $A$ and storing it with the memory footprint of $l^2$. As we can see from step 1), the input tensor has been loaded twice for two different projection operators; and, from step 3), the intermediate tensors $Q$ and $K$ have been stored and loaded without any further calculation, which is a waste of data movement. However, if we fused those operators together, which aligned with our definition of fused Op 1,  we will see the result will be the same but it will only cost one input tensor $X$ load and skip storing and loading the intermediate tensors $Q$ and $K$. This fusion will reduce the global memory $S3$ read and write by $5dl$ with a bit of shared scratchpad $S2$ pressure. We will explore the trade-off between the fusion scheme and the on-chip buffer size in the later.

\subsection{MSE: Mapping Space Explorer}
\label{subsec:mse}

As described in Algorithm \ref{alg:SAMT_kernel_fusion}, for each fusion scheme, we can get the $Mapping\_Candidates$ through \textbf{OFE} and MAESTRO\_FUSION with local scratchpad size $S1$, shared scratchpad size $S2$, NoCs bandwidth $B$, number of PEs $P$, transformer model matrix dimensions $M_i$, $N_i$, $K_i$ and $FusedOp$ (comes from \textbf{OFE}). Then for each $Mapping\_Candidates$, we perform a genetic search algorithm. As illustrated in Fig. \ref{fig:framework}, after initializing the population of the mapping candidates, we perform $Crossover$, $Mutation$, and $Reorder$. $Crossover$: it will randomly select two genomes (mappings) from the parent pool and interchange the tile size for those two genomes; $Mutation$: it will randomly generate a new dimension to parallelize if accelerator types allow and a new tile size for one dimension of the mappings; $Reorder$: it will swap the tile size of two dimensions from two mappings randomly. After evolution, MAESTRO\_FUSION will provide a performance analysis (latency, energy, PE utilization, etc.) and a fitness function will be applied based on the target optimization (latency, energy, $S1$ size, $S2$ size, etc.). Here we picked latency as our first optimization goal and energy as the second. The selection process will keep track of the current best mapping strategy. After selection, we update the parents pool and elites pool before entering the next iteration.

We enable two types of outputs here: fixed-dataflow type and flexible dataflow type. Fixed dataflow means the same dataflow mapping except the tiling sizes will be applied to each operator in the entire model; flexible dataflow means different dataflow mappings will be enabled for different operators in the transformer model.
\begin{figure}[ht]
    \centering 
    \includegraphics[width=\columnwidth]{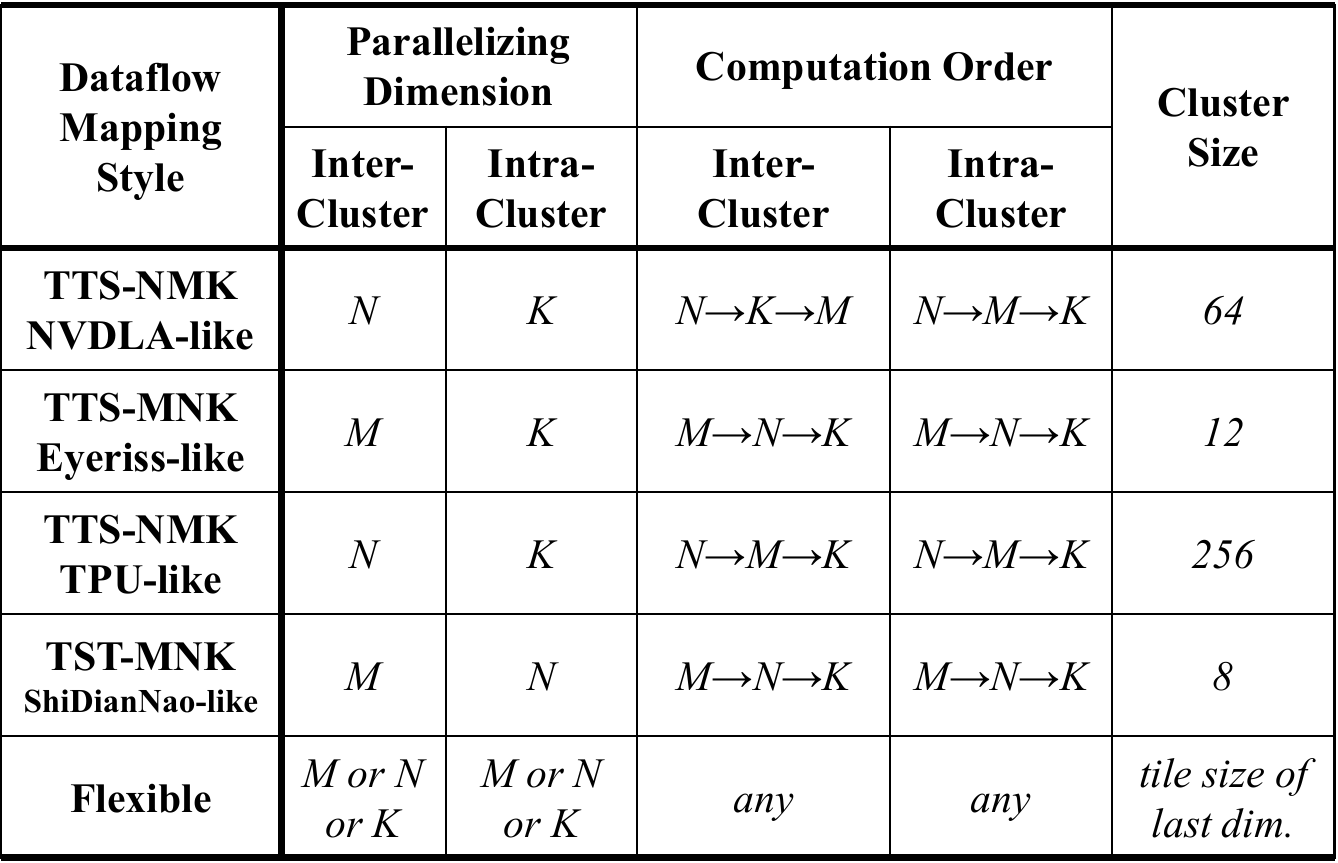}
    \caption{Dataflow mapping styles of different types of spatial accelerators}
    \vspace{-1mm}
    \label{fig:acc_type}
\end{figure}

\renewcommand{\arraystretch}{1.2}
\begin{table*}[]
\centering
\caption{Definition of Fused Operators and Corresponding Memory Footprints}
\begin{adjustbox}{width=520pt}
\begin{tabular}{ccccccccc|c}
\hline 
ID        & \textbf{Fused Op}  & \textbf{Orignal Ops}                                                    & \textbf{\begin{tabular}[c]{@{}c@{}}Memory \\ Fused\end{tabular}} & \textbf{\begin{tabular}[c]{@{}c@{}}Input Tensor \\ Fused\end{tabular}} & \textbf{\begin{tabular}[c]{@{}c@{}}Output \\ Tensor \\ Fused\end{tabular}} & \textbf{\begin{tabular}[c]{@{}c@{}}Memory \\ Orginal\end{tabular}} & \textbf{\begin{tabular}[c]{@{}c@{}}Input Tensor \\ Original\end{tabular}} & \textbf{\begin{tabular}[c]{@{}c@{}}Output Tensor \\ Original\end{tabular}} & \textbf{\begin{tabular}[c]{@{}c@{}}Memory \\ Reduced\end{tabular}} \\[0.2cm] \hline
\textbf{1} & $A=W_Q\otimes X\otimes W_K\otimes X$          & \begin{tabular}[c]{@{}c@{}} $Q=W_Q\otimes X$, \\ $K=W_K\otimes X$, \\ $A=Q\otimes K$ \end{tabular}     & $2d^2+l^2+dl$                                                    & $2d^2+dl$                                                              & $l^2$                                                                        & $2d^2+l^2+6dl$                                                    & $2d^2+4dl$                                                               & $l^2+2dl$                                                                  & \boldsymbol{$5dl$}                                                     \\[0.5cm]
\textbf{2} & $S=Softmax(Q\otimes K)$     & \begin{tabular}[c]{@{}c@{}}$A=Q\otimes K$,\\ $S=Softmax(A)$\end{tabular}           & $l^2+2dl$                                                        & $2dl$                                                                  & $l^2$                                                                        & $3l^2+2dl$                                                        & $l^2+2dl$                                                                 & $2l^2$                                                                      & \boldsymbol{$2l^2$}                                                     \\[0.5cm]
\textbf{3} & $O=V\otimes Softmax(A)$     & \begin{tabular}[c]{@{}c@{}}$S=Softmax(A)$,\\ $O=V\otimes S$\end{tabular}           & $l^2+2dl$                                                        & $l^2+dl$                                                                & $dl$                                                                        & $3l^2+2dl$                                                        & $2l^2+dl$                                                                 & $l^2+dl$                                                                    & \boldsymbol{$2l^2$}                                                     \\[0.5cm]
\textbf{4} & $O=(W_V\otimes X)\otimes S$         & \begin{tabular}[c]{@{}c@{}}$V=W_V\otimes X$,\\ $O=V \otimes  S$\end{tabular}                 & $d^2+l^2+2dl$                                                    & $d^2+l^2+dl$                                                            & $dl$                                                                        & $d^2+l^2+4dl$                                                      & $d^2+2dl+l^2$                                                            & $2dl$                                                                      & \boldsymbol{$2dl$}                                                     \\[0.5cm]
\textbf{5} & $Y=(W_0\otimes (V \otimes S)) $         & \begin{tabular}[c]{@{}c@{}}$O=V \otimes  S$,\\ $Y=W_0\otimes O$\end{tabular}                 & $d^2+l^2+2dl$                                                    & $d^2+l^2+dl$                                                            & $dl$                                                                        & $d^2+l^2+4dl$                                                      & $d^2+2dl+l^2$                                                            & $2dl$                                                                      & \boldsymbol{$2dl$}                                                     \\[0.5cm]
\textbf{6} & $F=a_2\otimes GELU(a1\otimes Y+b_1)+b_2$ & \begin{tabular}[c]{@{}c@{}}$L_1=GELU(a_1\otimes Y+b_1)$,\\ $L_2=GELU(a_2\otimes L_1+b_2)$\end{tabular} & $2((d_{FFN}+l)d)$                                                  & $2d_{FFN}d+dl$                                                          & $dl$                                                                        & $2(d_{FFN}(l+d)+dl)$                                                & $d_{FFN}d+dl$                                                               & $d_{FFN}l$                                                                    & \boldsymbol{$2d_{FFN}l$}                                                 \\[0.2cm] \hline
\end{tabular}
\end{adjustbox}
\begin{tablenotes}
   \item  *$l$ : Sequence Length, *$d$ : Embedding Size,  *$d_{FFN}$ : Hidden Dimension of Linear Projection Layer  
\end{tablenotes}
\label{tab:op_fuse}
\end{table*}



\begin{algorithm}
    \footnotesize
    \caption{SAMT Kernel Fusion Algorithm}
    \label{alg:SAMT_kernel_fusion}
    \begin{algorithmic}[1]
        \STATE Input: Matrix dimensions for each layer of the transformer model: $M_i$, $N_i$, $K_i$; The local scratchpad size: $S1$; The shared scratchpad size $S2$; The bandwidth for NoC: $B$; The total number of PEs: $P$; The type of accelerator: $Arch$, Generation for search: $\alpha$; $Target1$ and $Target2$ the optimization goal (chosen from latency, energy, $S1$ size, $S2$ size, $P$).
        \STATE Output: Datacentric Mapping descriptions for the optimal mapping strategy
        \STATE $Dataflow\_Candidates$ = Get\_Dataflow( $Arch$ )
        \FOR{$Fusion\_id = 1 \to Num\_Funsion\_Scheme$}
            \STATE $FusedOp$ = Get\_Fused\_Operators($Fusion\_id$)
            \STATE $Mapping\_Candidates$ = MAESTRO\_FUSION ( $S1$, $S2$, $B$, $P$, $M_i$, $N_i$, $K_i$, $FusedOp$ )
            \FOR{$i = 1 \to Dataflow\_Candidates$}
\WHILE{$Current\_Iteration$ \textless $\alpha$}
                    \STATE ( $P.D.$, $C.O.$, $T.S $ ) = Crossover( $Mapping\_Candidates$ )
	              \STATE ( $P.D.$, $T.S.$ ) = Mutation( $Mapping\_Candidates$ )
	              \STATE ( $ C.O $ ) = Reorder( $Mapping\_Candidates$ )
                    \STATE Update\_Childs()
                    \IF{$Childs\_Fitness$($Target1$, $Target2$) \textgreater $Threshold$}
                        \STATE Update\_Elites()
                        \STATE Update\_Parents()
                    \ENDIF
                \ENDWHILE
            \ENDFOR
        \ENDFOR
    \end{algorithmic}
\end{algorithm}


\subsection{MAESTRO\_FUSION: Cost Model}

After enabling the dynamic fusion scheme for the transformer model, we need to evaluate the performance of the spatial accelerators to find the optimal dataflow mapping. We tried to leverage the detailed analytical modeling framework, MAESTRO \cite{MaestroJun}, to describe regular transformer operators and our proposed fused operators in \textbf{OFE}. However,  the original MAESTRO does not support fused operators from \textbf{OFE}. Thus, we expanded the native MAESTRO to \textbf{MAESTRO\_FUSION} to enable dataflow directives given a fused operator. \textbf{MAESTR\_FUSION} extended fused operator support by converting the S2/DRAM access to more efficient inter-PE communication and can report the latency of the given model, energy usage of the hardware accelerator, and memory access count on $S1$ scratchpad and $S2$ scratchpad given the hardware configurations. The efficiency of \textbf{MAESTR\_FUSION} lies in its swift analysis of diverse forms of data reuse within the accelerator. This comprehensive set of metrics offers valuable insights into the performance characteristics of the spatial accelerator. Inputs to \textbf{MAESTR\_FUSION} are made of a transformer model description and hardware resource details, provided in the form of lists, while the mapping is articulated through a data-centric representation. This representation encompasses three directives, facilitating a succinct and compiler-friendly depiction of mappings.

\subsection{How different mapping affects data reuse}

\begin{figure}[ht]
    \centering 
    \includegraphics[width=\columnwidth]{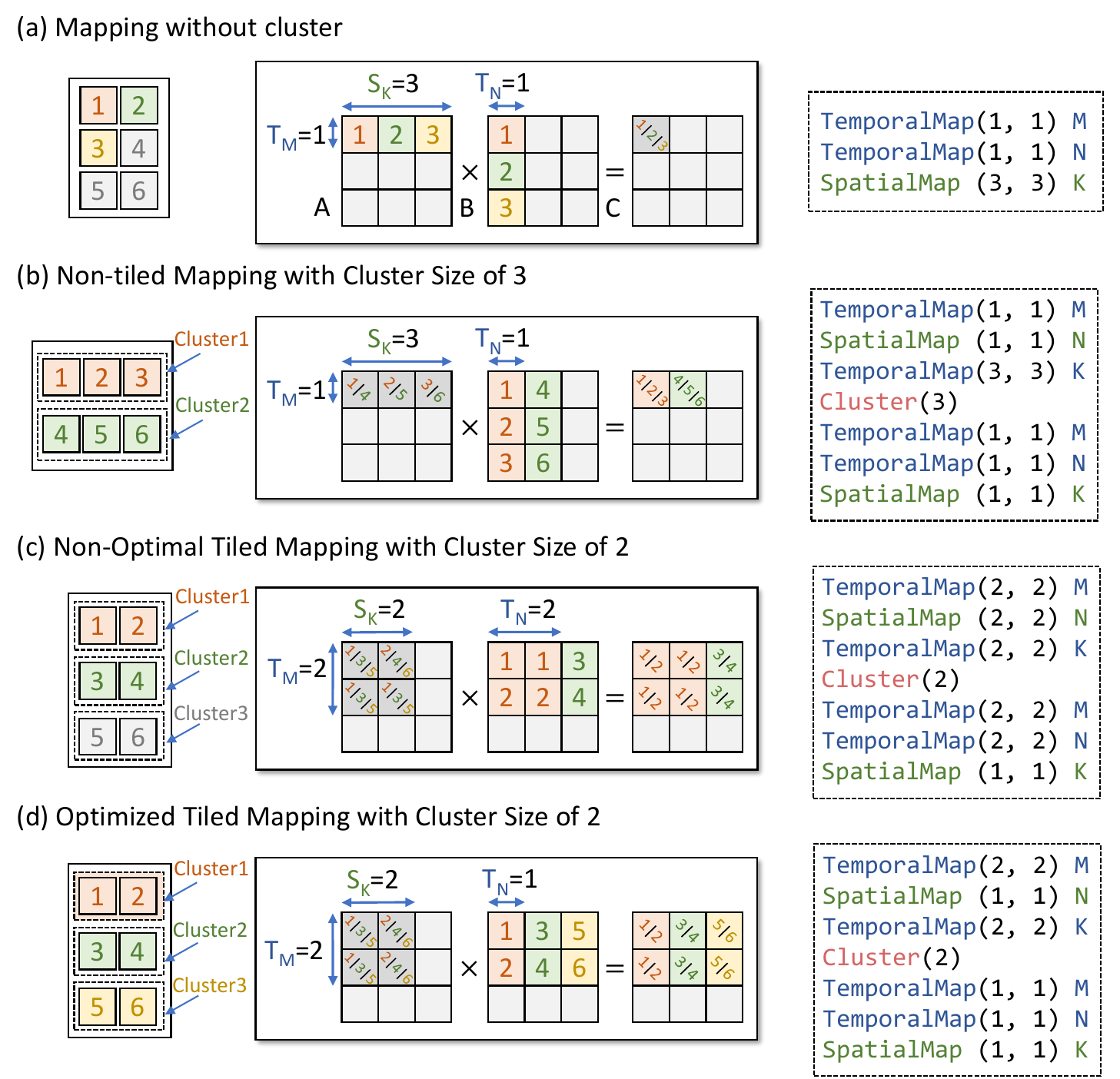}
    \caption{Different Mapping Strategies}
    \label{fig: different mappings}
\end{figure}

For each mapping, which refers to the specific dataflow strategy, we can describe it using computation order, parallelism, and tiling strategy. Each mapping determines how much data will be used at each PE at the time of processing. These data need to be moved from the off-chip memory to the shared scratchpad $S2$, and from the shared scratchpad $S2$ to the local scratchpad $S1$, as shown in Fig. \ref{fig:spatial arch}. Data reuse happens when some portion of data remains in the local scratchpad $S1$ or shared scratchpad $S2$. Operator fusion further increases the chance to keep some of the data in on-chip memory rather than sending it back to the off-chip memory. 

For a mapping, if the dimension of the matrix is larger than the physical dimension of the PE array, then the matrix must be tiled. As shown in Fig. \ref{fig: different mappings}, it shows the impact of different tile sizes for mapping a GEMM (a $3 \times 3$ matrix $A$ multiplied with a $3 \times 3$ matrix $B$) on a spatial architecture with 6 PEs. We use $T_M$, $T_N$ and $S_K$ to represent the temporal mapping for $M$ and $N$ dimension and spatial mapping for $K$ dimension. As computation order and parallelizing dimension are fixed in both inter-cluster and intra-cluster in this example, different cluster sizes and tiling sizes will affect the data reuse opportunity. In Fig. \ref{fig: different mappings} (a), a non-cluster mapping has been applied here. Only 3 PEs (number 1 to 3) are spatially mapped onto the K dimension, which is the column of matrix $A$ and the row of the matrix $B$. With this mapping, the entire accelerator is computing one entry of the matrix $C$ and leaving 3 PEs (from number 3 to 6) idle. In Fig. \ref{fig: different mappings} (b), we fully map 3 PEs from each cluster onto the $K$ dimension. All 6 PEs are utilized to compute two entries of matrix $C$. However, this non-tiled mapping does not provide the optimal performance. In Fig. \ref{fig: different mappings} (c), we split 6 PEs into 3 clusters. 2 PEs from each cluster are spatially mapped onto $K$ dimension and yet fully cover the $K$ dimension, which is called the tiled-mapping. However, this time we temporally map the $M$ and $N$ dimensions with size 2, which means each PE will compute two different data across the time. This will yield an underutilization of cluster 3 because there is not enough data in the $N$ dimension to allocate for cluster 3. Finally, Fig. \ref{fig: different mappings} (d) shows an optimized tiled mapping strategy with a cluster size of 2. All 2 PEs in each cluster will be fully utilized and this enables maximum data reuse.

\subsection{Hardware design space and dataflow mapping space}

To enable flexible spatial accelerator design, our framework defines four key configurations in Algorithm \ref{alg:SAMT_kernel_fusion} for the hardware specification: number of PEs $P$, shared scratchpad size $S2$, local scratchpad size $S1$, and network-on-chip (NoC) bandwidth $B$. Even though it is impossible to do an apple-to-apple comparison with previous accelerator works, we try to normalize the hardware configurations across different works for a fair comparison. We evaluated all the different mapping strategies previous works can provide, and assign the same hardware configurations to all the accelerators under comparison. For example, in GEMM, the $K$ dimension is reduced when calculating the output matrix, thus we spatially map the K dimension within the cluster of PEs for the accelerators that support spatial reduction (TPU-like, Eyeriss-like, NVDLA-like). These accelerators support spatial reduction through forward-and-store or reduction trees. However, the ShiDianNao-style accelerators, do not support spatial reduction, thus we spatially map the N dimension within the cluster instead of the K dimension. Listed in Fig. \ref{fig:acc_type}, we summarized the important mapping information of different types of accelerators. For example, $TTS-NMK$ represents that intracluster mapping style is temporally map $N$ and $K$ dimensions and spatially map $K$ dimension, which aligns with fixed dataflow NVDLA-like accelerators. The parallelizing dimension, computation order, and cluster size for those accelerators are fixed except for the flexible one. Thus, given the type of accelerator $Arch$ (e.g., NVDLA-like, TPU-like, Fleixble), we can determine the data-centric directives for the mapping space of a transformer model.

\begin{figure*}[ht]
    \centering 
    \includegraphics[width=\linewidth]{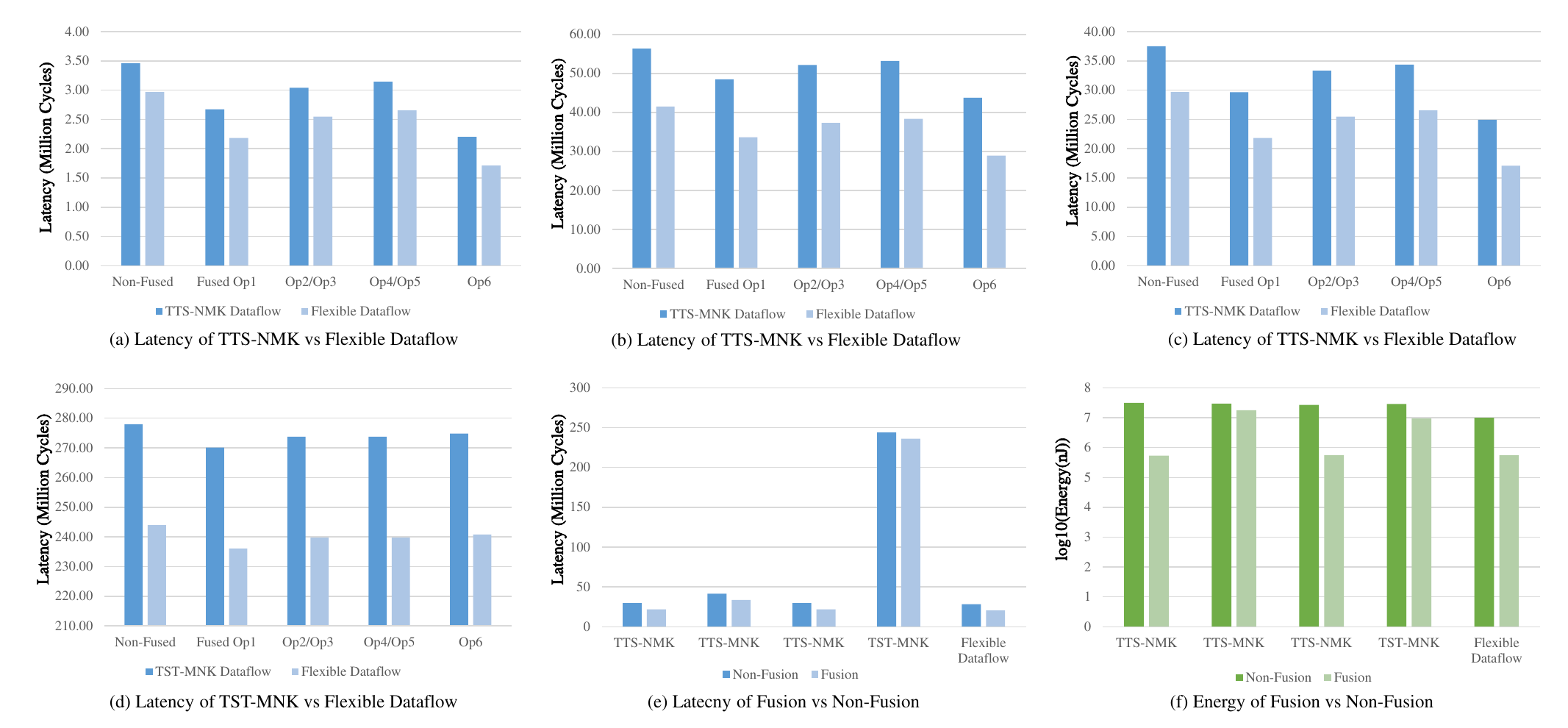}
    \caption{Evaluation Results of Latency and Energy Consumption under Various Settings}
    \label{fig: latency_op}
\end{figure*}

\section{Evaluation}
\label{sec:evaluation}
\begin{table}[]
\centering
\caption{Example of different Hardware Configurations}
\begin{adjustbox}{width=0.48\textwidth}
\begin{tabular}{c|c|c|c|c|c}
\hline
\textbf{Config} & \textbf{\# of PEs} & \begin{tabular}[c]{@{}c@{}}\textbf{S1} \\ \textbf{Size}\end{tabular} & \begin{tabular}[c]{@{}c@{}}\textbf{S2} \\ \textbf{Size}\end{tabular} & \begin{tabular}[c]{@{}c@{}}\textbf{NoC} \\ \textbf{Bandwidth}\end{tabular} & \begin{tabular}[c]{@{}c@{}}\textbf{Off-Chip} \\ \textbf{Mem. Bandwidth}\end{tabular} \\ \hline
Edge \cite{Coral}          & 256      & 256B                                                     & 20 MB                                                   & 16  GB/s                                                 & 80 GB/s (DRAM)                                                     \\ \hline
Mobile \cite{NPU}        &4098      & 512B  
  & 40 MB                                                   & 40 GB/s
                                     & 80 GB/s (DRAM)
                        \\ \hline
Cloud \cite{jouppi2023tpu}        & 65536    & 2048B                                                     & 100 MB                                                  & 800 GB/s                                                 & 1000 GB/s (HBM)                                                    \\ \hline
\end{tabular}
\end{adjustbox}
\label{tab:HW_config}
\end{table}


\begin{table}[]
\centering
\caption{Latency and Energy Reduction under Various S2 Scratchpad Sizes}
\begin{tabular}{cccc}
\toprule
\begin{tabular}[c]{@{}c@{}}\textbf{S2 Size} \\ \textbf{(MB)} \end{tabular} & \textbf{Fusion Code} & \begin{tabular}[c]{@{}c@{}}\textbf{Latency Reduced} \\ \textbf{(Million Cycles)}\end{tabular} & \begin{tabular}[c]{@{}c@{}}\textbf{Energy Reduced} \\ \textbf{($\mu$J)}\end{tabular} \\
\midrule
12 & 110101 & 24.64 & 52.74 \\ \hline
15 & 110101 & 27.79 & 59.47 \\ \hline
17 & 111101 & 31.98 & 68.45 \\ \hline
20 & 111111 & 35.13 & 75.18 \\
\bottomrule
\end{tabular}
\label{tab:reduction_s2_size}
\end{table}


\begin{figure}
    \centering
    \includegraphics[width=1.0\linewidth]{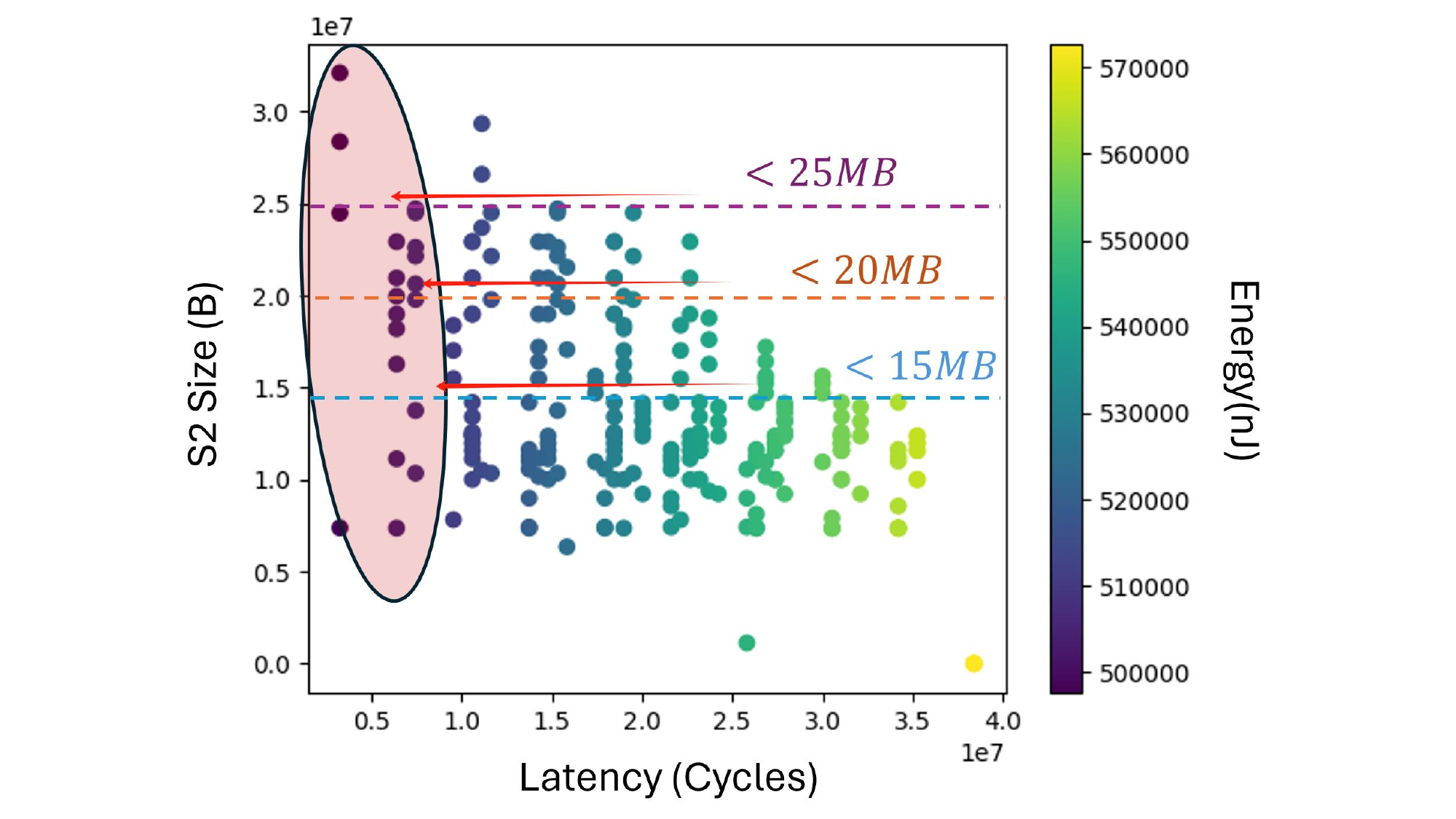}
    \caption{The Pareto-Front Solutions Found by \Design}
    \label{fig:S2vslatencyvsenergy}
\end{figure}

In this section, we show that \Design framework can find Pareto-front solutions given different Transformer models and spatial accelerators.

We evaluated five types of accelerators as shown in Fig. \ref{fig:acc_type}. For consistent comparison, each accelerator was configured with identical hardware settings (refer to the Edge setting in Table \ref{tab:HW_config}), including 256 Processing Elements (PEs), 256 Bytes for S1, 20 MB for S2, and a Network on Chip (NoC) bandwidth of 16 GB/s. We used the GPT-2 transformer model in this analysis, which has an embedding size of 768 and a sequence input length of 1024). Fig. \ref{fig: latency_op} (a)-(d) demonstrates how different accelerator designs and the integration of operator fusion techniques influence latency improvements. Specifically, Fig. \ref{fig: latency_op} (a) highlights that the TTS-NMK accelerators (as per the setup in Fig. \ref{fig:acc_type}) experience a 14\% latency reduction with a flexible dataflow approach without fusion, and between 15\% and 22\% with basic fusion primitives. Further, Fig. \ref{fig: latency_op} (b) to (d) indicate latency improvements ranging from 12\% to 26\% using flexible dataflow without fusion, and from 13\% to 34\% with basic fusion strategies. Finally, Fig. \ref{fig: latency_op}(e) and (f) show that under the same hardware configurations, flexible dataflow combined with optimal fusion can lead to up to 91\% latency reduction and up to 23\% energy savings compared to fixed dataflow without fusion.
\begin{figure}
    \centering
    \includegraphics[width=1.0\linewidth]{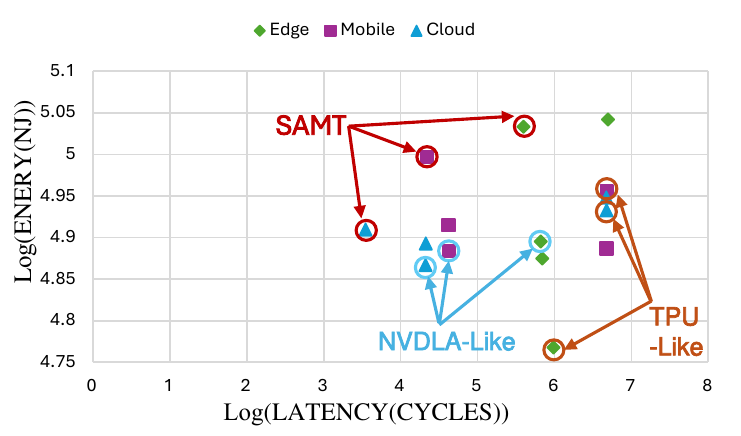}
    \caption{Latency vs. Energy of different dataflow accelerators across different platforms}
    \label{fig:dse}
\end{figure}

TABLE \ref{tab:reduction_s2_size} shows the trade-off between the latency and energy reduction because of operator fusion and $S2$ scratchpad size. As $S2$ size increases, we can incorporate a more aggressive fusion scheme, which means more operators will be fused and that requires a larger $S2$ capacity. In short, larger $S2$ further reduces the runtime latency and energy consumption. 

Fig. \ref{fig:S2vslatencyvsenergy} shows the Pareto-front solution sets found by \Design, where each point represents a fusion scheme supported and enabled by OFE. This figure illustrates the strong correlation between latency and energy within out hardware setting. Given the hardware constraints, especially the size of $S2$, we observe various fusion schemes could lead to optimal latency reduction, which aligns well with the findings from TABLE \ref{tab:reduction_s2_size}. We notice that, within $S2$ limits, for example, $S2$ less than 25 MB, the different fusion schemes along the dashed line demonstrate the potential to reduce latency without increasing energy consumption.

Fig. \ref{fig:dse} shows the optimal fixed dataflow and flexible dataflow accelerators with fusion reported by \Design in the edge, mobile, and cloud settings as detailed in Table.\ref{tab:HW_config}. In each of those settings, \Design consistently identifies solutions that achieve the lowest latency while maintaining relatively low energy consumption, which aligns with our prioritization of latency first and energy second. The figure highlights the advantages of employing flexible dataflow accelerators compared to fixed dataflow accelerators (NVDLA-like and TPU-like) when \Design can successfully determine the optimal dataflow mapping for them with the fusion. 

To further show the potential of flexible dataflow accelerators in \Design for transformer model, we consider the different computation patterns for the prefilling and decoding stage of GPT-3. The prefill stage of the decoder-only model (GPT-3) has the same computation pattern as the encoder-only model (BERT-Base). However, the decoding stage of the decoder-only model has an autoregressive computation pattern. From the \Design report (assuming GPT-3 Medium model with a prompt length of 128, and sequence length of 1024), the latency and energy for each iteration using the same dataflow as the prefill stage are $2.5 \times 10^{10}$ cycles and $5.1 \times 10^8$ nJ, different from using optimized flexible dataflow, whose latency and energy are $1.8 \times 10^8$ cycles and $6.6 \times 10^7$ nJ. It shows the potential of flexible dataflow accelerators in decoder-only transformer models.

\section{Related Work}
\label{sec:relatedwork}
Previous research has addressed deep learning accelerator loop ordering and mapping strategies, primarily through compiler optimizations. These efforts typically target multiple platforms and favor architectures with broader applicability. While some of these works incorporate dataflow mappings as one aspect of optimization, they often do not explicitly explore the design of dedicated specialized accelerators, as we do in our current study.

Additionally, there are works that specifically focus on optimizing dataflow mapping and exploring the design space for accelerators. However, many of these efforts lack considerations at the operator fusion level. Moreover, the majority of them are tailored for Convolutional Neural Networks (CNNs), whereas our research explicitly targets transformer networks, incorporating detailed analysis of the heavily leveraged operator fusion scheme.
\subsection{Compiler Optimization}

TVM \cite{chen2018tvm} is an open-source machine learning compiler framework that automates the process of optimizing neural network models for a variety of hardware targets, including CPUs, GPUs, and specialized accelerators. It uses a deep learning-based approach to determine efficient operator fusion, optimizing the performance of the computational graph by considering the cache size and memory hierarchy of the target hardware. Although TVM provides extensive support for multiple hardware platforms, its optimization strategies focus mainly on traditional computing units and do not intrinsically consider the unique needs of emerging spatial architectures for Transformer models. Unlike \textbf{SAMT}, TVM does not explicitly adjust its strategy based on the different data flows and hardware configurations required by space accelerators (such as the number of PEs, NoC architecture, and memory bandwidth).

XLA \cite{xla} is another compiler that can convert high-level representations of neural networks into optimized machine code for CPUs, GPUs, and other ML accelerators. XLA's strength lies in its ability to tightly integrate with computation graph, enabling significant performance boosts. However, XLA's optimizations are generally confined to CPU and GPU architectures with existing LLVM backend support. For deployment on non-CPU-like hardware, XLA requires extensive backend development, which can be a resource-intensive process. This limitation is particularly relevant when considering spatial architectures for which XLA does not naturally provide support, contrasting sharply with \textbf{SAMT}’s direct focus on optimizing dataflow mappings specific to such accelerators.

Ansor \cite{ansor} uses a cost model to guide its search for the optimal kernel implementation across various hardware platforms. However, Ansor still primarily targets traditional accelerator designs and does not address the complexities associated with spatial architectures that are central to \textbf{SAMT}. Ansor’s general approach is to optimize operator-level performance without specific consideration for the architectural variations and dataflow strategies that are essential for the next-generation hardware accelerators used in processing Transformer models.

\subsection{Accelerator Mappings}
Kao et al. \cite{gamma} introduced the GAMMA framework, designed to automate the hardware mapping of Deep Neural Network (DNN) models onto accelerators. By employing Genetic Algorithms, they sought to optimize the mapping process, thereby enhancing the efficiency and performance of DNN accelerators. The utilization of genetic algorithms involves a heuristic search strategy inspired by natural selection processes, leading to iterative refinement of hardware mappings. It is noteworthy, however, that GAMMA primarily focuses on mapping strategies for Convolutional Neural Networks (CNNs), and its applicability to mainstream transformer models is limited due to distinct parallelization and data reuse schemes inherent to these models.

MAERI \cite{MAERI} focuses on the evaluation of spacial accelerator architectures that employ a tiled general matrix-matrix multiplication (GEMM) kernel. The study introduces a framework dedicated to identifying optimized mappings, including dataflow and tile sizes, for a tiled GEMM tailored to a given spatial accelerator and workload combination. The evaluations, conducted across five spatial accelerators, demonstrate that the systematically generated tiled GEMM mappings outperform various GEMM workloads on diverse accelerators, emphasizing the potential for high performance achieved through thoughtful optimization.

The FLAT framework, as introduced by Kao et al. \cite{kao2023flat}, presents a dataflow optimization strategy addressing attention bottlenecks in neural network models. Attention mechanisms, integral for capturing intricate data relationships, often introduce operational intensity and other bottlenecks on existing hardware. FLAT employs a tiling technique and limited operator fusion to augment data reuse opportunities, thereby reducing memory requirements and increasing arithmetic intensity.

On the other hand, our \textbf{SAMT} framework further refines the operator fusion scheme. It facilitates the co-optimization between various transformer models and hardware configurations, ultimately determining the optimal fixed or flexible (if hardware-supported) dataflow mapping.

\section{Conclusion}

In this work, we shed a light to the unexplored design space in the intersection of operator fusion and flexible dataflow for addressing low arithmetic intensity challenges in Transformer models. We explored all the possible operator fusion in Transformer blocks enabled by recent integer arithmetic algorithms on non-trivial operators such as softmax and presented benefits from the new operator fusion space.

We codified our methodologies into a framework \Design that consists of operator fusion explorer \textit{\textbf{OFE}}, an extended MAESTRO cost-model supporting operator fusion \textit{\textbf{MAESTRO\_FUSION}}, and genetic algorithm-based mapping space explorer \textit{\textbf{MSE}}. Our framework not only leverages the dynamic operator fusion schemes for the transformer models but also co-search the optimal dataflow mapping strategies for spatial accelerators. Our evaluation shows that such a comprehensive approach leads to significant latency and energy benefits: reduce inference latency by 12\% to 91\% and energy consumption by 3\% to 23\%. Such promising results motivate future works to consider the extended fusion and dataflow mapping optimization for Transformer accelerators, which can be built on top of our \Design framework.


\bibliographystyle{IEEEtranS}
\bibliography{refs}

\end{document}